\documentclass[10pt,conference]{IEEEtran}
\IEEEoverridecommandlockouts
\usepackage[utf8]{inputenc}
\usepackage[T1]{fontenc}
\usepackage{amsmath,amssymb,amsfonts}
\usepackage{mathtools}
\usepackage{dsfont}
\usepackage{algorithmic}
\usepackage{graphicx}
\usepackage{textcomp}
\usepackage[linesnumbered]{algorithm2e}
\usepackage{xcolor}
\usepackage{microtype}
\usepackage[backend=biber,
bibstyle=ieee,
citestyle=numeric-comp,
sortcites=true,                             
maxbibnames=3
]{biblatex}
\usepackage{ifthen}

\makeatletter
\newcounter{IEEE@bibentries}
\renewcommand\IEEEtriggeratref[1]{%
  \renewbibmacro{finentry}{%
    \stepcounter{IEEE@bibentries}%
    \ifthenelse{\equal{\value{IEEE@bibentries}}{#1}}
    {\finentry\@IEEEtriggercmd}
    {\finentry}%
  }%
}
\makeatother
\usepackage[margin=10pt,font=small,labelfont=bf]{caption}
\usepackage{subcaption}
\bibliography{refs}
\usepackage[nonumberlist]{glossaries}
\newacronym{gral}{GRAL}{Graph-based Localization}
\newacronym{wsn}{WSN}{Wireless Sensor Network}
\newacronym{mwsn}{MWSN}{Mobile Wireless Sensor Network}
\newacronym{vslam}{VSLAM}{visual simultaneous localization and mapping}
\newacronym{tdoa}{TDoA}{time difference of arrival}
\newacronym{tof}{TOF}{time of flight}
\newacronym{rssi}{RSSI}{radio signal strength indicator}
\newacronym{apit}{APIT}{approximate point in triangle}

\usepackage{amsthm}

\begin{document}

\title{GRAL: Localization of Floating Wireless Sensors in Pipe Networks\\
}

\author{\IEEEauthorblockN{Martin Haug}
\IEEEauthorblockA{\textit{Technische Universität Berlin}\\
Berlin, Germany \\
m.haug@tu-berlin.de}
\and
\IEEEauthorblockN{Felix Lorenz\textsuperscript{*}}
\IEEEauthorblockA{\textit{Technische Universität Berlin}\\
Berlin, Germany \\
felix.lorenz@tu-berlin.de}
\and
\IEEEauthorblockN{Lauritz Thamsen}
\IEEEauthorblockA{\textit{Technische Universität Berlin}\\
Berlin, Germany \\
lauritz.thamsen@tu-berlin.de}
}

\maketitle
\begingroup\renewcommand\thefootnote{*}
\footnotetext{Work done while at Technische Universität Berlin, now at ecospace}
\endgroup

\begin{abstract}
    Mobile wireless sensors are increasingly recognized as a valuable tool for monitoring critical infrastructures.
    An important use case is the discovery of leaks and inflows in pipe networks using a swarm of floating sensor nodes.
    While passively drifting along, the devices must track their individual positions so critical points can later be located.
    Since pipelines are often situated in inaccessible places, large portions of the network can be shielded from radio and satellite signals, rendering conventional positioning systems ineffective.
    
    In this paper, we propose a novel algorithm for assigning location estimates to recorded measurements once the sensor node leaves the inaccessible area and transmits them via a gateway.
    The solution is range-free and makes use of a priori information about the target pipeline network.
    We further describe two extended variants of our algorithm which use data of encounters with other sensor nodes to improve accuracy.
    Finally, we evaluate all variants with respect to various network topologies and different numbers of mobile nodes in a simulation.
    The results show that our algorithm localizes measurements with an average accuracy between 4.81\% and 7.58\%, depending on the variability of flow speed and the sparsity of reference points.
\end{abstract}

\section{Introduction}\label{sec:introduction}

In order to operate and maintain a pipe network, utility operators must be able to investigate its current state and detect specific undesirable conditions such as leakages or inflows/infiltrations.
However, with pipelines buried deep underground, direct access and large-scale deployment of stationary sensors for in vivo monitoring is often infeasible.
There have been recent efforts to resolve this using autonomous mobile robots~\cite{nassiraei2007,tuatar2007} and~\glsfirst{mwsn}~\cite{lai2010,kim2010,sun2011}.
A~\gls{mwsn} is a network of nodes that can move, detect changes in their environment and communicate wirelessly.
In the special case of pipe monitoring, nodes can do entirely without active locomotion and instead float with the current of the contained liquid~\cite{ishihara2012}.
Thus, they can be produced at a fraction of the cost of actively moving robots and are at the same time more robust. 
Sensor nodes can be inserted into the system at likely points of interest. They may be recovered or disposed of at a wastewater treatment plant.  
For a comprehensive overview of the literature on~\gls{mwsn} for pipe inspection, see~\cite{bensaleh2013}.\\

The use of a big fleet of sensors to monitor a pipe network requires solutions for data collection and scalable analytics \cite{lorenz_scalable2020, geldenhuys_dependable2021}.
However, to make use of the recorded data in the first place, the position in which the measurement was taken must be known.
Satellite-based localization systems like GPS are a natural choice for localization but are often not applicable in~\gls{mwsn} due to power constraints or insufficient signal coverage.
In such cases, one can instead make use of information on the proximity to other nodes and the properties of the environment to estimate the position of each reading.

\begin{figure}
  \centering
    \includegraphics[width=0.98\columnwidth]{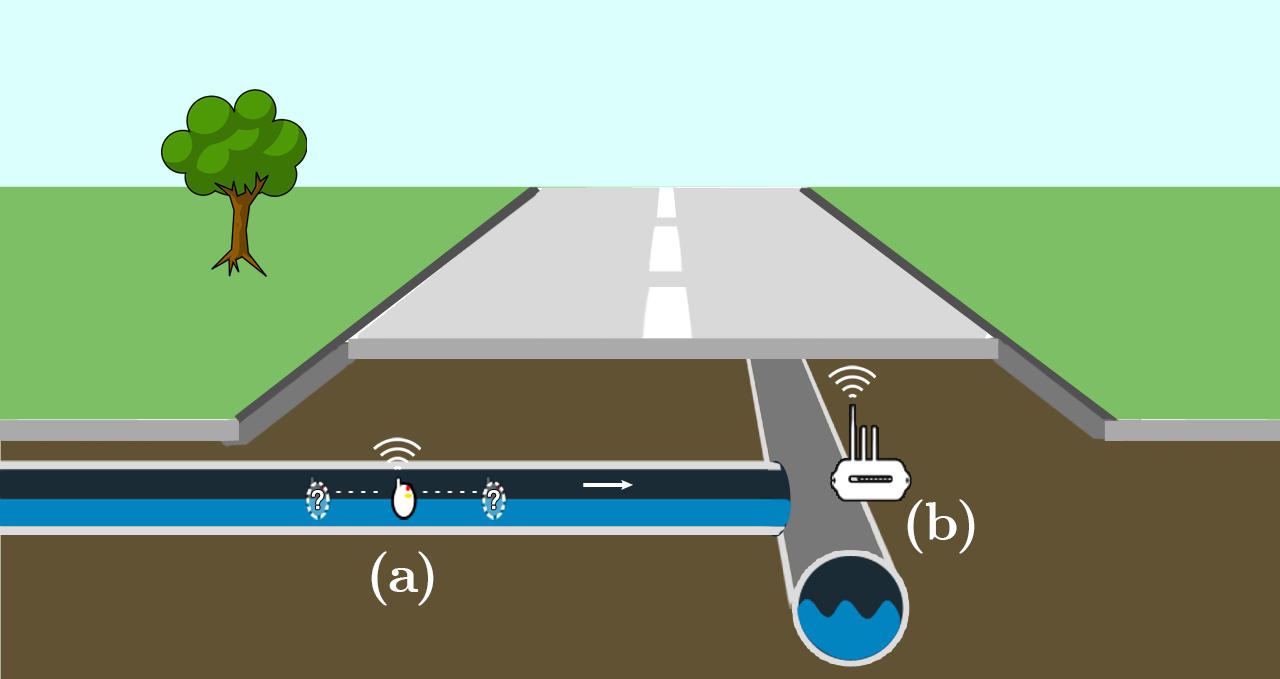}
    \caption{The measurements taken by  a sensor node (a) are submitted to the GRAL backend through a gateway (b) which resides at a junction in the pipe network.}
    \label{fig:usecase}
\end{figure}

In this paper, we address the localization problem of~\glspl{mwsn} deployed in wastewater networks.
However, the approach can also be used in other pipeline-like environments like water networks or oil pipelines.
We formalize the localization problem in Section~\ref{sec:problemsetup}.
We review existing localization algorithms for both~\glspl{wsn} in general and \glspl{mwsn} with floating nodes in Section~\ref{sec:relatedwork}.
For our work, we assume that the environment is tree-shaped, where nodes move along the edges from the leaves to the root.
Gateways are installed at specific points throughout the pipe network.
We introduce the~\gls{gral} algorithm as a method for sensor localization in such situations (Section~\ref{sec:algorithm}).
The idea is to use~\gls{mwsn} gateways as anchors and interpolate the path between them, where nodes traverse areas without coverage.
Reconstruction of measurement locations is done at a central backend, to which a node sends its recordings upon entering the radio range of a gateway.
We analyze the performance of~\gls{gral} with several pipe network topologies using a simulation (Section~\ref{sec:results}).

\section{Problem Setup: Localization in Pipe Networks}\label{sec:problemsetup}

Applications face the challenge of tracking the location of a sensor node while it is moving through a pipe networks.
For our solution to this, we consider networks that satisfy the following requirements:
\begin{itemize}
    \item The topology of the network can be described as
    \begin{enumerate}
        \item \emph{Junctions} that may feature stationary gateways with uplink connections  
        \item \emph{Links} of a known length connecting the junctions
    \end{enumerate}
    \item This topology is fixed and can be modeled as a tree-shaped graph in which the vertices correspond to junctions and the edges represent links
    \item The stationary gateways' approximate wireless radius and positions are known a priori
    \item Sensor units record relevant physical quantities such as conductivity or temperature while floating through the pipes
    \item Every time a sensor unit passes a gateway, it will send a batch of measurements to the backend
    \item The liquid flow direction in the network is never reversed. The sensor nodes thus always travel from the leaves to the root
\end{itemize}

This leads us to the~\emph{measurement localization problem} that is addressed in this paper:

\emph{Given a series of measurements from a mobile wireless sensor node floating through a tree-shaped pipe network, determine the location in that network at which each reading was taken.}

\section{Related Work}\label{sec:relatedwork}
Classical approaches to~\gls{wsn} localization in arbitrary environments can be divided into two classes:
\emph{Range-based} localization schemes use observations of \emph{anchors} with known positions to estimate distances whereas \emph{range-free} algorithms which only rely on the binary state of connectivity~\cite{mao_wireless_2007}.
\emph{Lateration} is a classic example for a range-based localization method.
It relies on the~\gls{rssi}, together with a model for radio wave propagation~\cite{rappaport_wireless_2002}.
Another range-based method relies on the~\emph{time difference of arrival} between a radio pulse and an ultrasonic pulse to estimate the distance to the anchors~\cite{krishnamachari_networking_2005}.
Most range-based approaches are not easily applicable for usage in ~\gls{mwsn} due to the high number of anchors required, poor calibration of cheap radio devices, and the presence of obstacles~\cite{karl_protocols_2007}.

In networks with dense anchor placement, reasonable results can already be achieved by simply using the centroid of all visible reference nodes~\cite{singh_range_2015}, a simple range-free approach.
A graph-based solution is described in~\cite{priyantha_anchor-free_2003}, where a fold-free graph embedding is found, upon which mass-spring based optimization is applied to localize a node.
Again, the problem with respect to the applicability of classical range-free methods in pipe networks is that a high gateway density can rarely be guaranteed in practice.

Both kinds of approaches have been used in MWSN monitoring with floating sensors: Range-free localization can be accomplished by applying RFID beacons to the outside of pipes~\cite{almazyad2014}.
The SewerSnort system uses a custom model of radio propagation in pipes to provide ranged localization with fewer, if more expensive, anchors~\cite{kim2009}.
Pipe networks allow for less conventional ranged approaches using the propagation characteristics of ultrasonic sound~\cite{bando2016} or electromagnetic waves~\cite{seco2016}, which, however, require dedicated hardware.

Another idea is to use~\gls{vslam}, i.e., rely on camera images of the pipes' interior to track the device's location~\cite{krys2007}.
Here, the main drawback is that the nodes either have to transmit large amounts of data or do expensive computations involved in~\gls{vslam}.

For a more comprehensive overview, consult~\cite{abbas2018}, a dedicated survey of localization solutions for pipe monitoring.
Another approach to localize leaks and inflows into water grids may be to forgo~\gls{wsn} entirely and to use methods like distributed temperature sensing or gas injection instead~\cite{adedeji2017}.

\section{The GRAL Algorithm}\label{sec:algorithm}

The graph-based localization algorithm solves the Measurement Localization Problem from Section \ref{sec:problemsetup} by assigning positions to the individual measurement packages. 
For this class of applications, the position of the nodes along the length of the pipes are of interest, therefore a position can be expressed as a four-tuple.
This tuple $\pi$ contains two junctions in the pipe network and the offset on the shortest path between them which qualifies a node's position.

Our algorithm is based on the assumption that we know the structure of the pipe network the nodes are deployed in and, at various moments, their positions.
The most obvious example, for the scenario laid out above, is that when the sensors receive the signal of a gateway with maximum strength, they are right below it.
GRAL exploits this series of moments with known positions: It localizes a time-series of node measurements (called packages) by partitioning them into ``epochs'', each of which have a known final position.
The Graph-based localization algorithm combines this information with prior knowledge of the topology of the deployment environment, encoded in a weighted, tree-shaped graph \emph{(environment graph)}.

The fact that the signal of the gateway may be received only within a certain radius provides another positional anchor:
We can tell that a node is at the boundaries of the circle this radius spans around the gateway the moment it stops or starts receiving the signal.
Note that this could yield more than one possible position.

Based on the information of previous packages, we can see at which positions within the graph a node has been before and select one position for the package.
Suppose that a node starts receiving a given gateway's signal, this gateway $g_2$ is located at a junction in the network where three pipes meet.
The node is then localized to be in the pipe that is on the shortest path between the last gateway $g_1$ it has been at and $g_2$, exactly $g_2$'s radius removed from the junction.

Below, we have to formalize the various types of epochs that are created in the application above to see when existing epochs learn about their final positions, can be localized and when new epochs need to be created.
We need to introduce some notation for this:
$J$ is the maximum index of the packages for an epoch, $s^j_i$ is the $j$th-strongest signal strength of a gateway in package $i$, and $g^j_i$ is the gateway with the $j$th-strongest signal in package $i$.
$G^1$, finally, is the set of gateways with the strongest signals for all packages of an epoch.

GRAL has three epoch types:

\begin{itemize}
\item $\nu$ epochs where no gateway's signal is received
\item $\alpha$ epochs in which the strength of the strongest received gateway signal increases with every package
\item $\omega$ epochs in which the strength of the strongest received gateway signal decreases with every package
\end{itemize}

At the most basic, a new package requires a new epoch if it does not fit the type condition for the latest existing epoch.
\begin{equation}\label{eq:prop}
    \tau(P)\coloneqq
    \begin{cases}
        \nu, & \text{if}~\forall~j = 1..J~~G_j = \emptyset\\
        \alpha, & \text{if}~\forall~j = 1..J~~s_j^1 > s_{j-1}^1 \land g_j^1 = g_{j-1}^1 \\
        \omega, & \text{if}~\forall~j = 1..J~~s_j^1 \leq s_{j-1}^1 \land g_j^1 = g_{j-1}^1
    \end{cases}
\end{equation}

\begin{figure}[h]
  \centering
    \includegraphics[width=0.5\columnwidth]{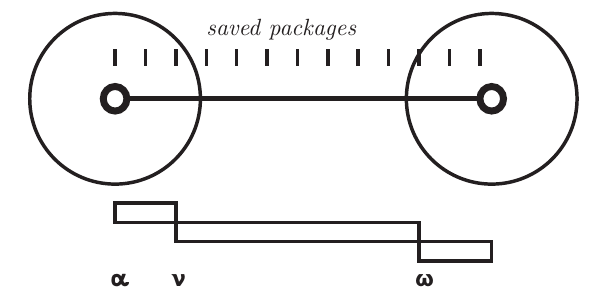}
    \caption{Epochs created for a single node moving from a gateway to another}
    \label{fig:epochs}
\end{figure}

\begin{algorithm}
    \SetKw{And}{and}
    \KwData{A node's package $p$ and a set of epochs $E$ for that node}
    \KwResult{The updated set of epochs $E$ annotated with the correct types $\tau$}
    \BlankLine
    $P_{\text{last}} \leftarrow P(\epsilon_{|E|})$ \tcp*{packages in last epoch}
    \tcc{determine package type as Eq.(\ref{eq:prop})}
    \eIf{$\tau(P_{\text{last}} \cup p) \in \{\alpha, \omega, \nu \}$}{
        $P_{\text{last}} \leftarrow P_{\text{last}} \cup p$\;
    }{
    \tcc{coalesce epochs if applicable}
        \eIf{$\exists i \in \mathbb{N} . \forall j \in \mathbb{N} .~j > i \land \tau(P_i) \neq \nu \wedge  \tau(P_j) = \nu$ \And $g^1(P_1(E_i)) = g^1(p)$}{
            coalesce $\{p\}$ and $\{ P(e) | e \in \epsilon_i \dots \epsilon_{|E|} \}$\;
        }{
        \tcc{otherwise, create new epoch}
            $E \leftarrow E \cup (\tau(\{p\}), $ \textemdash$, {p})$\;
        }
    }
	\Return{$E$}\;
    \BlankLine
    \caption{Integrating a new package with the epoch set, $P(i)$ returns the packages in the $i$th epoch}\label{alg:epochAlgorithm}
\end{algorithm}

Because GRAL interpolates the packages' positions within an epoch between a known starting and final position along the shortest route between those in the environment graph, both of these boundary positions have to be known so that the packages can be localized, i.e. the epoch is \emph{complete}.

There is a set of requirements of which an epoch must fulfill one to have a known final position $\pi_f$:

\begin{itemize}
\item It is of the type $\alpha$ and it is not the last epoch ($\pi_f$ is the position of $g^1_ {i}$)
\item It is of the type $\alpha$ and the RSSI of the gateway is maximal in its last package ($\pi_f$ is the position of $g^1_ {i}$)
\item There is a subsequent epoch that is not of the type $\nu$ ($\pi_f$ is the appropriate position at the border of $g^1_ {i+1}$)
\item The final position $\pi_f$ is pre-set
\end{itemize}

In order to be \emph{complete}, an epoch additionally needs a starting position which can either be pre-set or be derived from the final position of the preceding epoch (which must therefore also fulfill at least one of the above conditions).

For resiliency against positional fluctuations, all epochs at a specific gateway $g$ after the first $\alpha$ epoch are merged as long as no other gateway is seen.
GRAL is also fault-tolerant: If a gateway is not operable, the algorithm will just interpolate package positions with less epochs using the surrounding gateways.

We consider~\gls{gral} to be range-free since it does not use the~\gls{rssi} to measure a distance; instead, it uses preloaded data about the environment to derive distances.
The~\gls{rssi}, which the implementor can substitute for any other proximity indicator, is only used to determine the qualitative difference of increasing or decreasing proximity.

\subsection{Algorithm Extensions}
\label{sec:features}

In networks with sparse gateway deployment, epochs of the $\nu$ type containing many packages will be created.
GRAL is especially designed for use in non-pressurized wastewater systems with varying flow speeds.
Because interpolation error due to these speed differences accumulates over time, localization accuracy can be low in such environments.
To alleviate that issue, we propose two algorithmic extensions, which use information about nearby other sensor nodes to refine epochs improve position estimation.

\begin{figure}[h]
    \centering
    ~
    \begin{subfigure}[t]{0.4\columnwidth}
        \includegraphics[width=\textwidth]{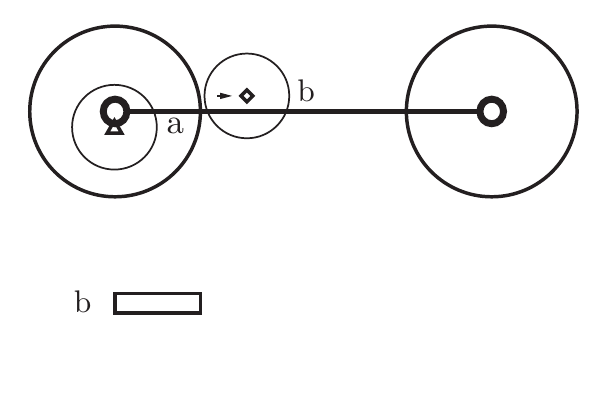}
        \caption{Node $b$ has already been at the left gateway, a second node $a$ arrives }
        \label{fig:check2}
    \end{subfigure}
    ~
    \begin{subfigure}[t]{0.4\columnwidth}
        \includegraphics[width=\textwidth]{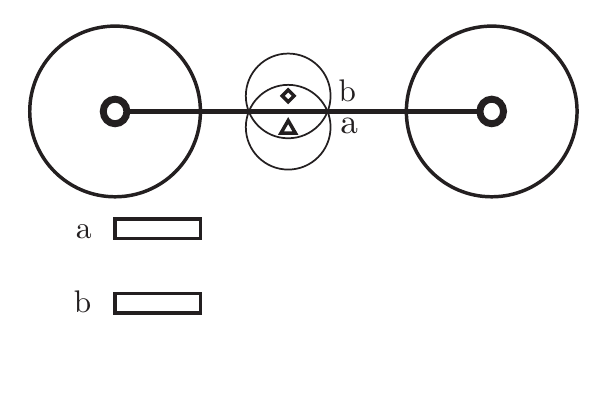}
        \caption{Nodes $a$ and $b$ encounter one another}
        \label{fig:check3}
    \end{subfigure}
    ~
    \begin{subfigure}[t]{0.4\columnwidth}
        \includegraphics[width=\textwidth]{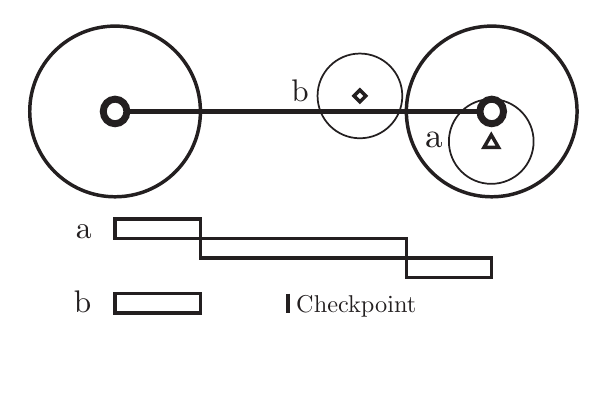}
        \caption{Node $a$ reaches the gateway, a checkpoint for $b$ is created}
        \label{fig:check4}
    \end{subfigure}
    ~
    \begin{subfigure}[t]{0.4\columnwidth}
        \includegraphics[width=\textwidth]{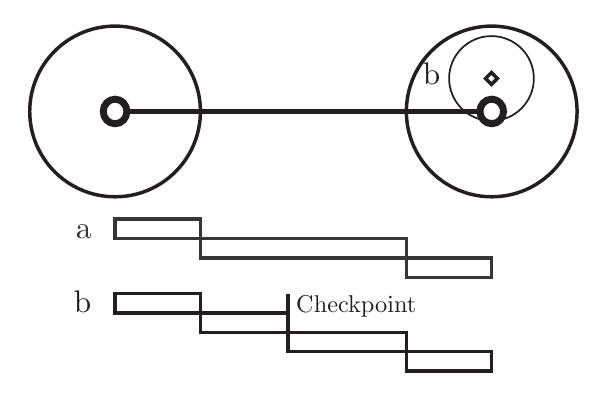}
        \caption{Node $b$ reaches the gateway, its $\nu$ epoch is split at the checkpoint}
        \label{fig:check5}
    \end{subfigure}
    \caption{Checkpoint creation on a single link}\label{fig:checkpoints}
\end{figure}

\textbf{Checkpointing}
If there is more than one mobile node in the system, the nodes may meet.
Data about these encounters can be used to make sure that both nodes get assigned a similar position for the moments where they have met.
Checkpoints are a way to accomplish this.
When processing an epoch's packages, a checkpoint containing a position, a timestamp and the identifier of the issuing node is added any encountered nodes.
A node can only add a single checkpoint to each other node per epoch -- the algorithm selects the wireless contact with the highest RSSI.
When the encountered node reaches a gateway and starts to transmit its packages, GRAL creates a new epoch for the packages with a timestamp later than that of the checkpoint.
The checkpoint's location is set as the position that was calculated for the node first to arrive at a gateway.
Fig.~\ref{fig:checkpoints} illustrates the process.

\textbf{Path rectification}
An encounter between two nodes can also sometimes place a lower bound on their positional estimates:
Imagine two nodes approaching a junction with three pipes, like in Fig.~\ref{fig:scenario-3}.
The nodes start on the left and had contact to respective left gateways before.
It is clear that, when the two nodes meet, they have to be situated behind the junction where they entered a shared path, because an encounter has not been possible on the disjunct paths they have been on before.

$\textrm{path}_v(v, w)$ returns the vertices on the path from $v$ (inclusive) to $w$ (exclusive) in the environment graph.

{
\setlength{\abovedisplayskip}{2pt}
\setlength{\belowdisplayskip}{2pt}
\setlength{\abovedisplayshortskip}{1.8pt}
\setlength{\belowdisplayshortskip}{1.8pt}
\begin{align}
    \exists v_c \in V . \neg \exists v_n \in V . \: & v_c \neq v_n \wedge v_c \in \textrm{path}_v(v_a, v_f)\, \wedge \nonumber \\
    &v_c \in \textrm{path}_v(v_b, v_f) \wedge v_n \in \textrm{path}_v(v_a, v_f)\, \wedge \nonumber \\
    &v_n \in \textrm{path}_v(v_b, v_f)\, \wedge \nonumber \\
    &|\textrm{path}_v(v_c, v_f)| < |\textrm{path}_v(v_n, v_f)| \label{eq:confluence}
\end{align}
\captionof*{figure}{
    If packages for a node $a$ coming from the position corresponding to vertex $v_a$ in the environment graph indicate that it has seen another node $b$ (coming from the corresponding position of vertex $v_b$), therefore, their \emph{confluence vertex} is calculated.
    }
}

The confluence vertex $v_c$ is as defined above as the earliest vertex that is contained both in the paths from $v_a$ and $v_b$ to a shared destination $v_f$.
Once the relevant epochs get completed, localization is first performed normally.
If the results of this process for $a$ estimate packages recording contact to $b$ to be located before the confluence vertex, the relevant epochs are split at the earliest package recording this contact with the location of the confluence vertex as the final position $\pi_f$ of the first of these two epochs, the packages are then re-localized.

\section{Evaluation}\label{sec:results}

To evaluate the performance of~\gls{gral} under varying conditions, we generate several pipe network graphs through which simulated sensor nodes move, sense, and communicate.
For this, we created a custom simulation environment.

\subsection{Simulation}

At each step of the simulation, all mobile node positions are moved a fixed amount towards the root junction, representing a base current within the pipes.
A noise term sampled from a boolean distribution with $P(1) = \frac{2}{3}$ is added to this movement.
Random fluctuations in flow speed are to be expected in real situations, due to obstructions and a non-uniform current along the cross-section of the pipe~\cite{haug_statistische_2006}. 
The simulation uses a fixed radio propagation model to calculate RSSI-like strength indicators for the connections. 
Just like a physical system, it occasionally emits batches of measurement packages to be localized by~\gls{gral}.
We compare our solution with and without the described extensions against a naive approach (baseline) which consists of simple linear interpolation between gateways based on the first- and last-contact timestamps.
A reference implementation of GRAL is available at \texttt{https://github.com/reknih/GRAL}.

\subsection{Experiments}

\begin{figure}[h]
    \centering
    \begin{subfigure}[t]{0.4\columnwidth}
        \includegraphics[width=\textwidth]{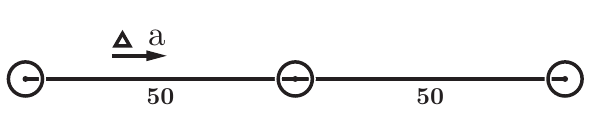}
        \caption{A single node in a pipe}
        \label{fig:scenario-1}
    \end{subfigure}
    ~
    \begin{subfigure}[t]{0.4\columnwidth}
        \includegraphics[width=\textwidth]{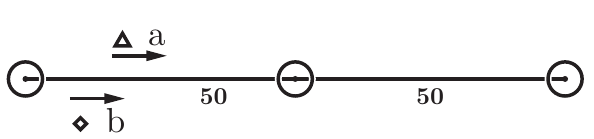}
        \caption{Two nodes in a pipe}
        \label{fig:scenario-2}
    \end{subfigure}
    \begin{subfigure}[t]{0.4\columnwidth}
        \includegraphics[width=\textwidth]{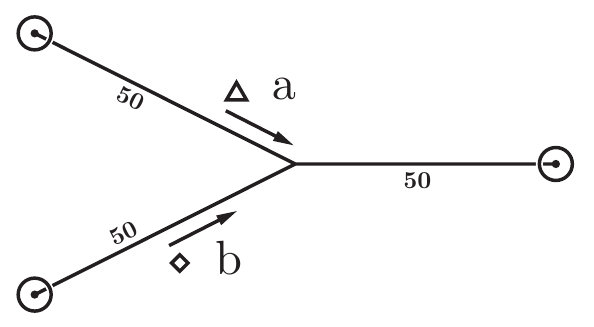}
        \caption{Two nodes in an environment with a confluence}
        \label{fig:scenario-3}
    \end{subfigure}
    ~
    \begin{subfigure}[t]{0.4\columnwidth}
        \includegraphics[width=\textwidth]{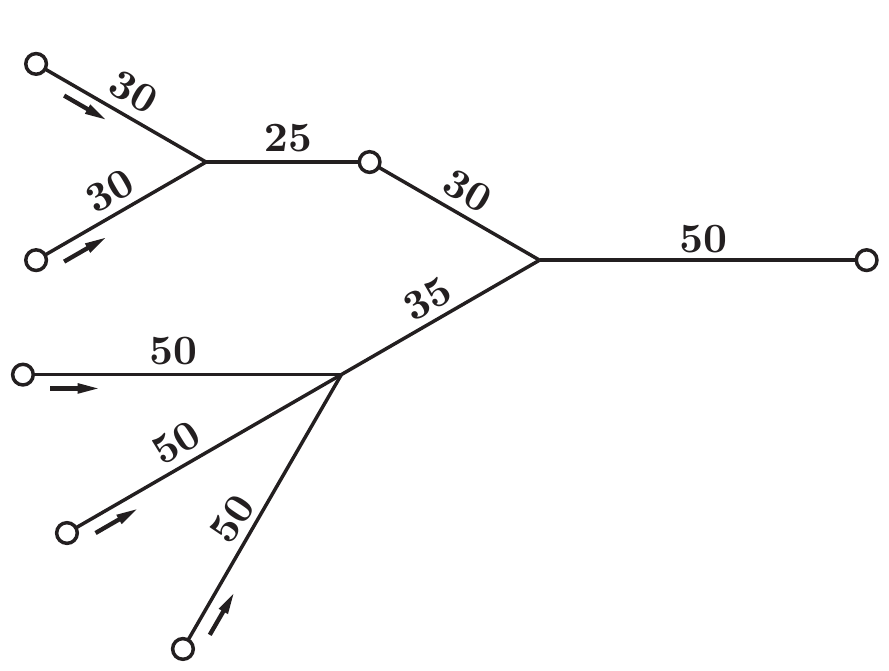}
        \caption{Complex environment}
        \label{fig:scenario-4}
    \end{subfigure}
    \caption{Overview of the test scenarios; to scale circles representing gateway ranges}\label{fig:test-scenarios}
\end{figure}

In total, our experiments cover four scenarios (overview in Fig.~\ref{fig:test-scenarios}) for which 200 randomized runs (\emph{instances}) are performed.
The test scenarios were constructed to be representative of possible topologies that sensor nodes might be deployed in in real wastewater systems.
Scenarios \ref{fig:scenario-3} and \ref{fig:scenario-4} include merging branches, which allow evaluation of the path rectification feature.
Deviations of the calculated locations from the ground truth are given as Root Mean Square Error for both each instance \emph{(iRMSE)} and the entire respective scenario \emph{(dRMSE)}.
We additionally use the Mean Absolute Error (MAE) to compute average error ranges in meters for real deployments.

\subsection{Results}

\begin{table}[htbp]
\caption{Evaluation results. Values correspond to RMSE over 200 instances. CP = Checkpointing, PR = Path rectification}
\label{tab:results}
\centering
\setlength{\tabcolsep}{6pt}
\begin{tabular}{llllll}
                                & Baseline                   & \multicolumn{4}{c}{GRAL}                                                                                      \\
                                &                            & \multicolumn{1}{c}{Vanilla} & \multicolumn{1}{c}{CP} & \multicolumn{1}{c}{PR} & \multicolumn{1}{c}{CP \& PR}  \\ \cline{2-6} 
\multicolumn{1}{l|}{Scenario 1} & \multicolumn{1}{l|}{9.54}  & \textbf{6.41}               & -                      & -                      & -                             \\
\multicolumn{1}{l|}{Scenario 2} & \multicolumn{1}{l|}{9.13}  & \textbf{6.5}                & 6.83                   & -                      & -                             \\
\multicolumn{1}{l|}{Scenario 3} & \multicolumn{1}{l|}{11.64} & 10.03                       & 9.9                    & 10.02                  & \textbf{9.89}                 \\
\multicolumn{1}{l|}{Scenario 4} & \multicolumn{1}{l|}{11.55} & 10.5                        & 11.72                  & \textbf{10.48}         & 11.71                         \\
\multicolumn{1}{l|}{Total}      & \multicolumn{1}{l|}{10.1}  & \textbf{8.36}               & -                      & -                      & -                            
\end{tabular}
\end{table}

The results of our experimental evaluation are summarized in Table~\ref{tab:results}.
\gls{gral} consistently scores a lower RMSE than the baseline.
In Scenario 3, the best accuracy is achieved by enabling both extensions, thus demonstrating the benefit of checkpoints in a relatively well-covered environment.
Throughout the following, we give a detailed discussion of the success and failure cases and project the results onto deployments in real pipe networks.\\

\textbf{Scenario 1} In the first scenario, the environment consists of three gateways, connected by two pipe segments and is populated with a single node (Fig.~\ref{fig:scenario-1}). 
\gls{gral} without extra features achieves a total error of $\textrm{dRMSE} = 6.41$, compared to the baseline error of $\textrm{dRMSE}_B = 9.54$. 
In this scenario, the MAE is $4.81$ over $100$ distance units.
This means that one can expect GRAL to have an average error of $4.81\%$ of the total route length for a localized point in this scenario, given that the ratio of pipe in range of a gateway equals $\frac{\sqrt{10}}{25}$.

\begin{figure}[h]
    \centering
    \begin{subfigure}[t]{0.37\columnwidth}
        \includegraphics[width=\textwidth]{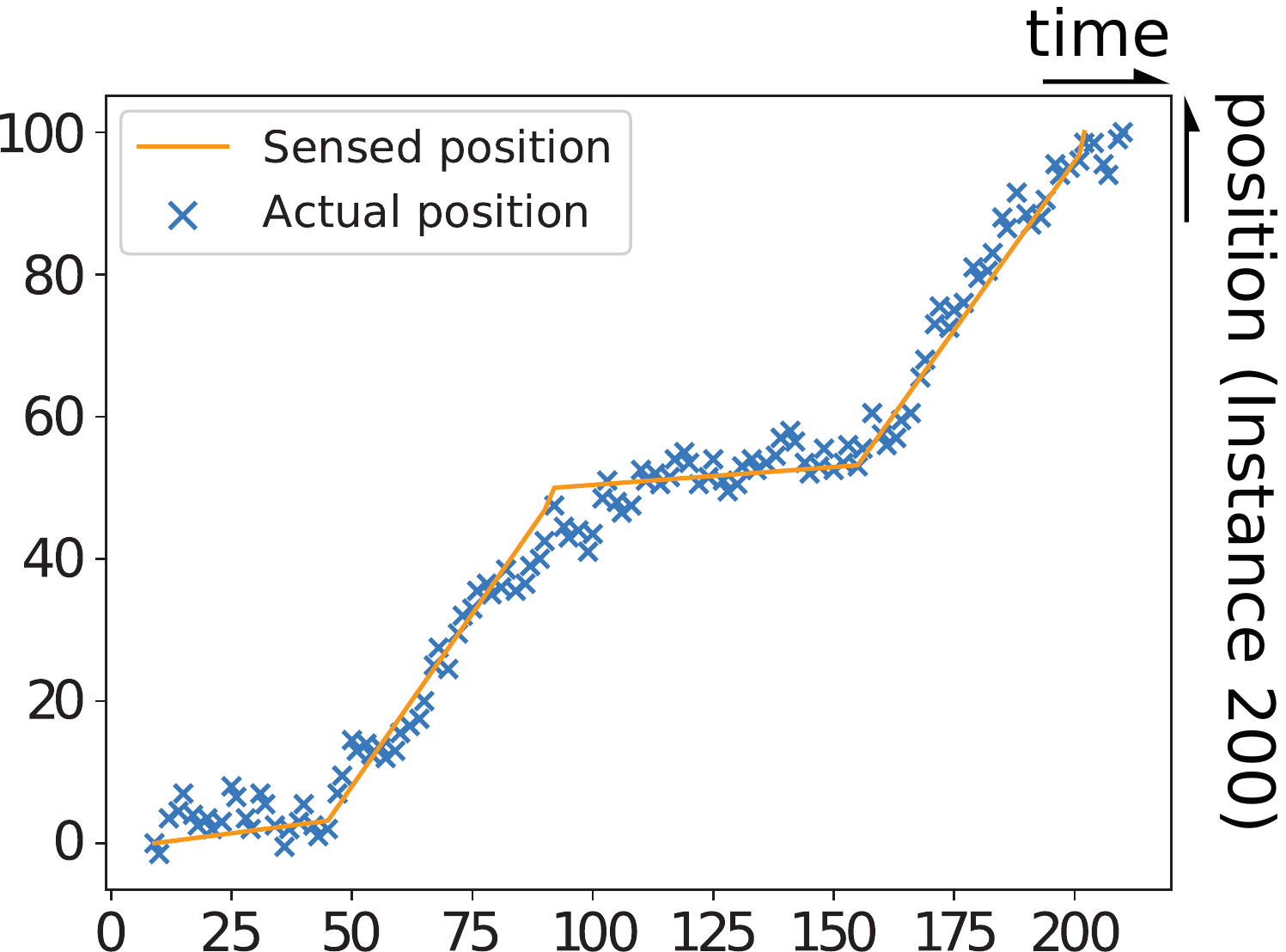}
        \caption{$\textrm{iRMSE} = 3.174$}
        \label{fig:good-i1}
    \end{subfigure}
    ~
    \begin{subfigure}[t]{0.37\columnwidth}
        \includegraphics[width=\textwidth]{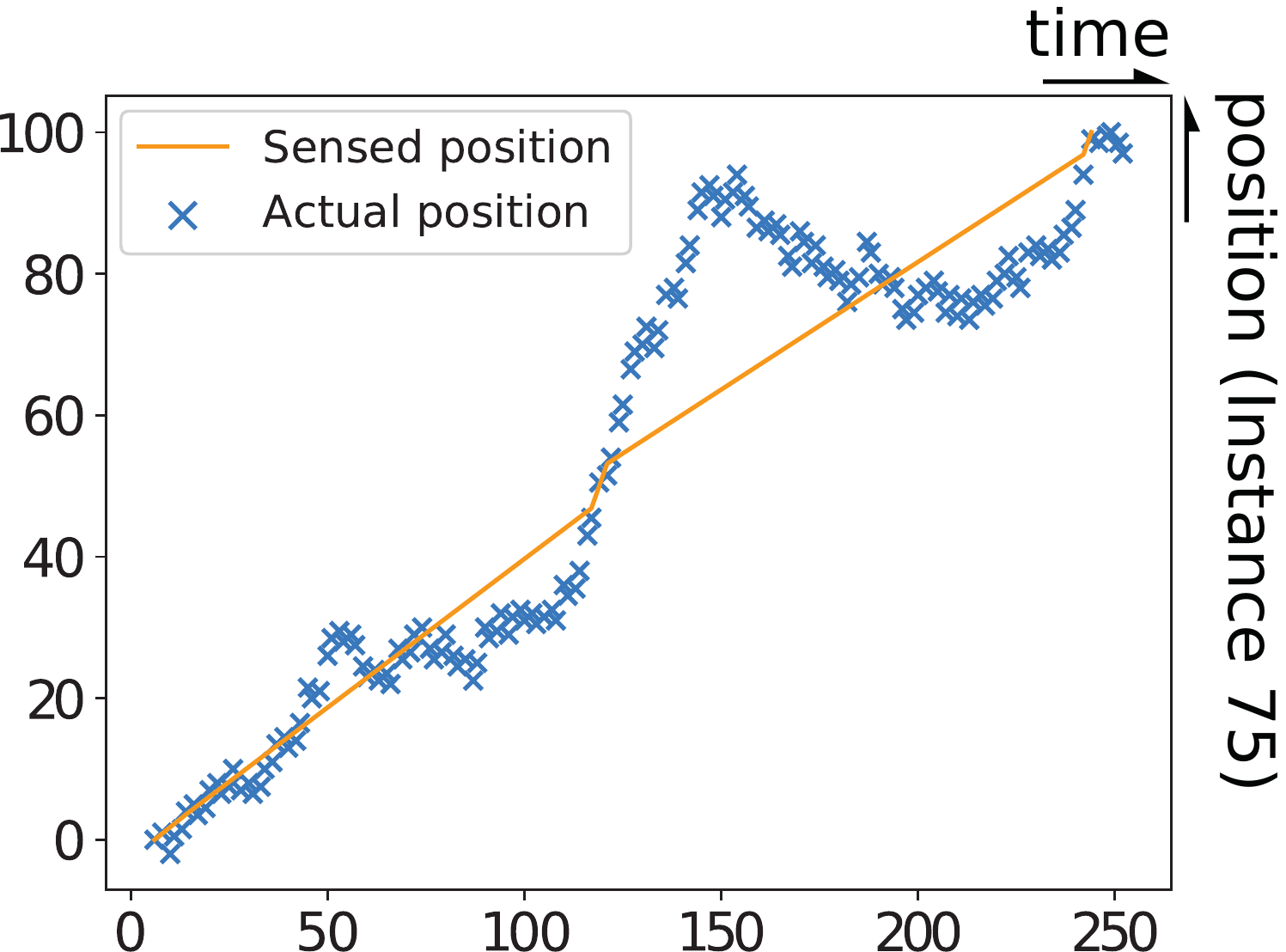}
        \caption{$\textrm{iRMSE} = 10.7$}
        \label{fig:bad-i1}
    \end{subfigure}
    \caption{Instances with low or high iRMSE from the first dataset}\label{fig:i1-examples}
\end{figure}

Fig.~\ref{fig:i1-examples} allows a closer look at instances with low and high errors in the first scenario, respectively.
In~\ref{fig:good-i1}, we see that the node briefly stays in the vicinity of the first and second gateway respectively, creating prolonged and flat $\omega$ epochs. 
In contrast, Instance~\ref{fig:bad-i1} shows the node to be drifting back and forth in the segment between two gateways, causing a large divergence from the interpolated values for that epoch.

\begin{figure}[h]
    \centering
    \includegraphics[width=0.45\columnwidth]{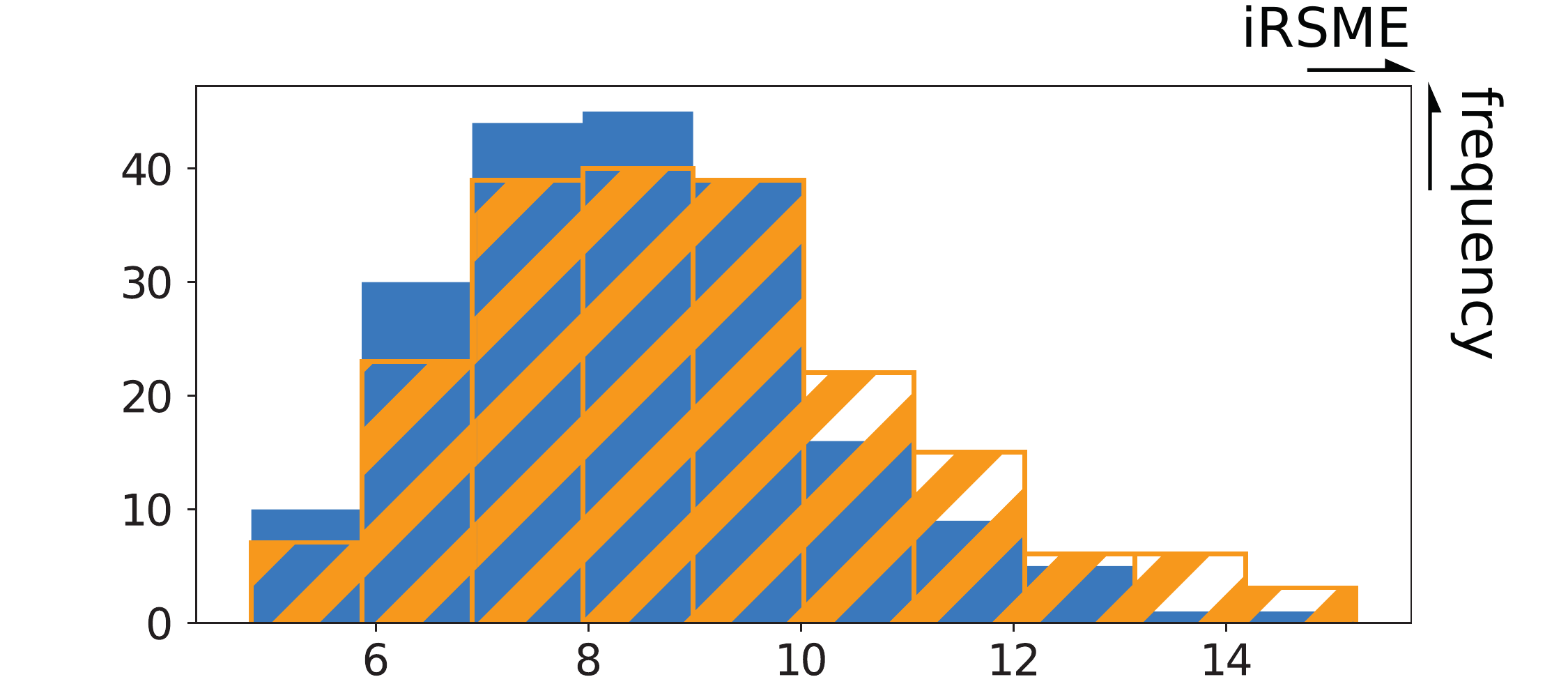}
    \caption{Distribution of iRMSE for a dataset with two nodes and no junctions; checkpoints enabled for striped bars}
    \label{fig:hist-i2}
\end{figure}

\textbf{Scenario 2} The next scenario as seen in Fig.~\ref{fig:scenario-2} has the same layout but with two nodes deployed in close succession.
Thus, the two are frequently within range of each other.
Fig.~\ref{fig:hist-i2} shows the distribution of iRMSE for localization with and without checkpoints respectively.
The total error for the former is $\textrm{dRMSE}_{C} = 6.83$, for the latter it is $\textrm{dRMSE} = 6.5$.
With the same assumptions as above, this would mean an average localization error of $5.22\%$ for checkpoints enabled and $4.98\%$ for checkpoints disabled.

\begin{figure}[h]
    \centering
    \begin{subfigure}[t]{0.37\columnwidth}
        \includegraphics[width=\textwidth]{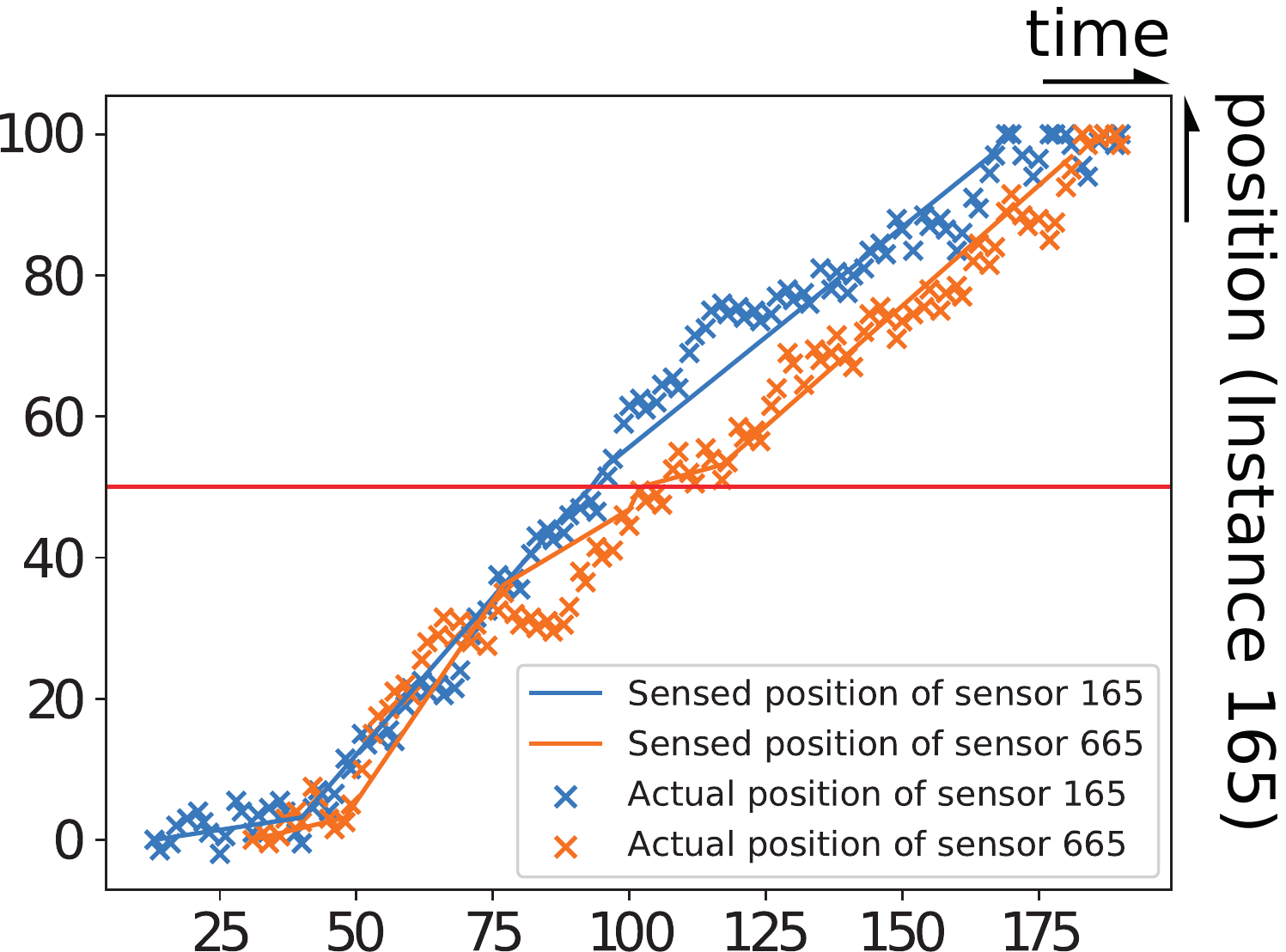}
        \caption{$\textrm{iRMSE} = 5.79$}
        \caption*{checkpoints enabled}
        \label{fig:good-chk-i2}
    \end{subfigure}
    ~
    \begin{subfigure}[t]{0.37\columnwidth}
        \includegraphics[width=\textwidth]{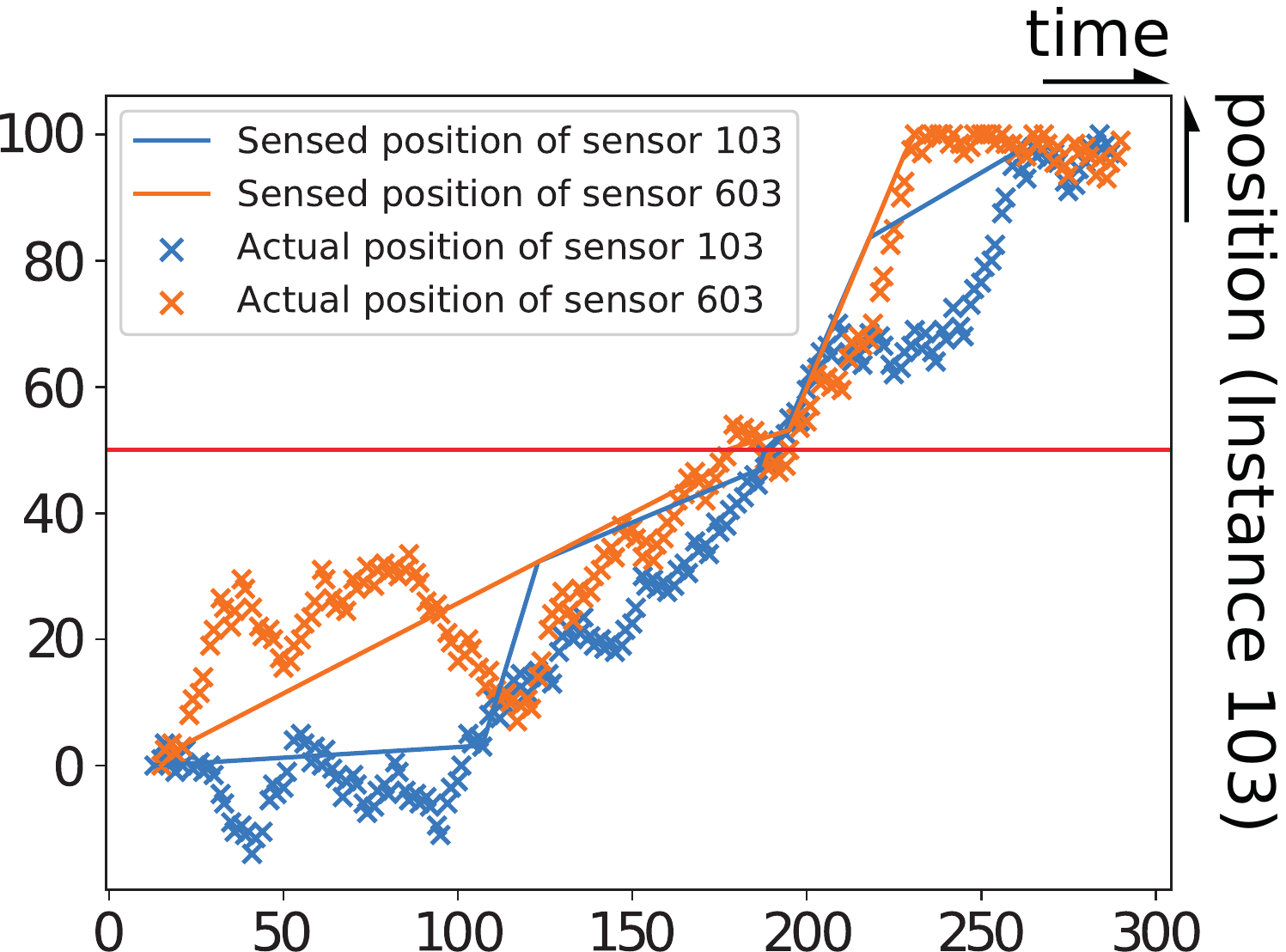}
        \caption{$\textrm{iRMSE} = 14.95$}
        \caption*{checkpoints enabled}
        \label{fig:bad-chk-i2}
    \end{subfigure}
    \begin{subfigure}[t]{0.37\columnwidth}
        \includegraphics[width=\textwidth]{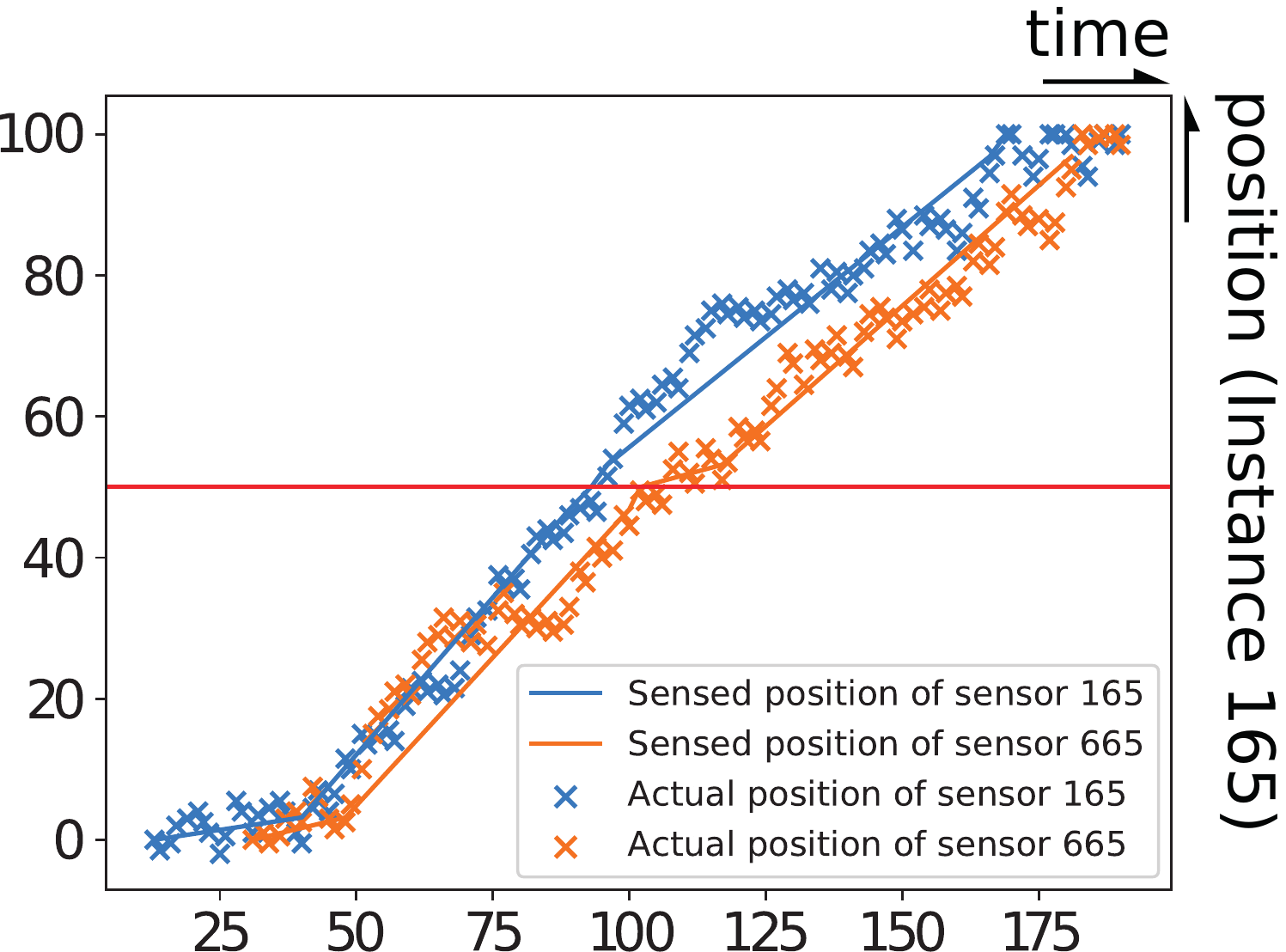}
        \caption{$\textrm{iRMSE} = 6.08$}
        \caption*{checkpoints disabled}
        \label{fig:good-none-i2}
    \end{subfigure}
    ~
    \begin{subfigure}[t]{0.37\columnwidth}
        \includegraphics[width=\textwidth]{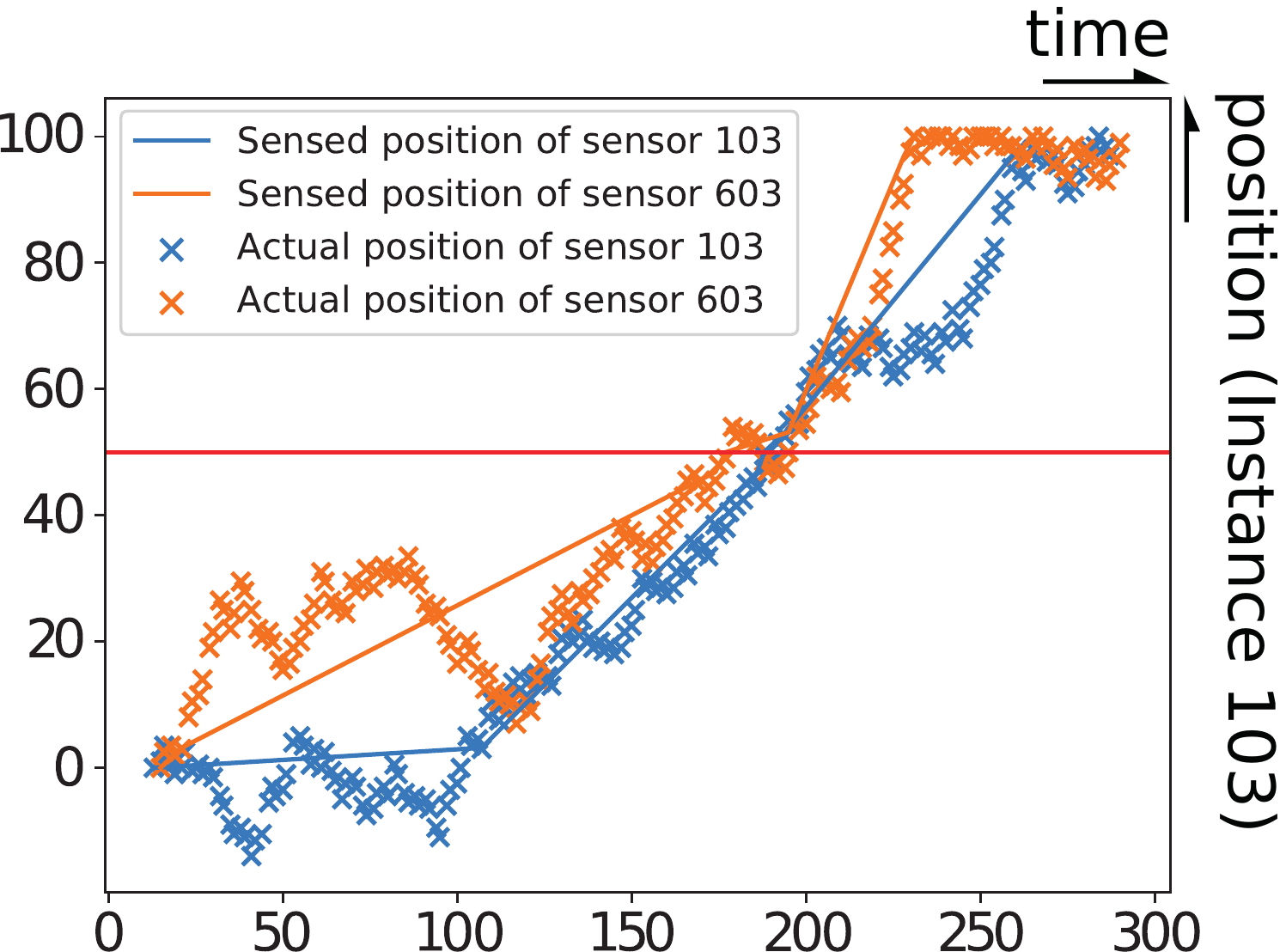}
        \caption{$\textrm{iRMSE} = 11.71$}
        \caption*{checkpoints disabled}
        \label{fig:bad-none-i2}
    \end{subfigure}
    \caption{Instances with low or high iRMSE from the second dataset}\label{fig:i2-examples}
\end{figure}

The decrease in performance with checkpointing enabled is a rather surprising result.
To see what has gone wrong, we again consider two exemplary instances in Fig.~\ref{fig:i2-examples}:
Node 165 (blue) in~\ref{fig:good-chk-i2} moves steadily, thus experiences small localization errors.
The speed of node 665 (orange) fluctuates between gateways.
Because node 165 is the first to arrive at a gateway, its relatively good localization can help improve the data for node 665.
Conversely, in \ref{fig:bad-none-i2} we again see one node (103; blue) moving steadily and one (603; orange) experiencing turbulence.
In this instance, the turbulent node 603 is the first to arrive at a gateway after it met its counterpart.
Thus, it passes on its big localization error to node 103.

We see here that checkpointing, while having a sound theoretical justification, can cause the propagation of errors from one node to another in certain cases.

\textbf{Scenario 3} This scenario introduces a new topology, as displayed in Fig.~\ref{fig:scenario-3}.
In this case, two initial gateways each serve as starting points for a node.
They are both connected to the same junction with pipes of equal length.
This center junction lacks a gateway but has another pipe leading to a last gateway.
In this scenario, path rectification becomes relevant as nodes may encounter other nodes that did not share all of their path before moving on to a shared gateway.
The results show that again vanilla~\gls{gral} ($\textrm{dRMSE} = 10.03$) outperforms the baseline algorithm ($\textrm{dRMSE}_B = 11.64$).
The total error is higher than in the previous datasets because of the uncertainty introduced by the missing center gateway. 
Interestingly, checkpoints seem to increase overall accuracy in this dataset ($\textrm{dRMSE}_C = 9.90$).
The total error decreases slightly with rectification enabled ($\textrm{dRMSE}_P = 10.02$).
Thus, simultaneously using both features, also increases localization accuracy ($\textrm{dRMSE}_{CP} = 9.89$).
This means, that if the fraction of each path covered by a gateway is $\frac{\sqrt{10}}{50}$, the mean localization error for the worst configuration (vanilla) is $7.62\%$.
The best configuration with checkpoints and path rectification would have an mean localization error of $7.58\%$.

\begin{figure}[h]
    \centering
    \begin{subfigure}[t]{0.37\columnwidth}
        \includegraphics[width=\textwidth]{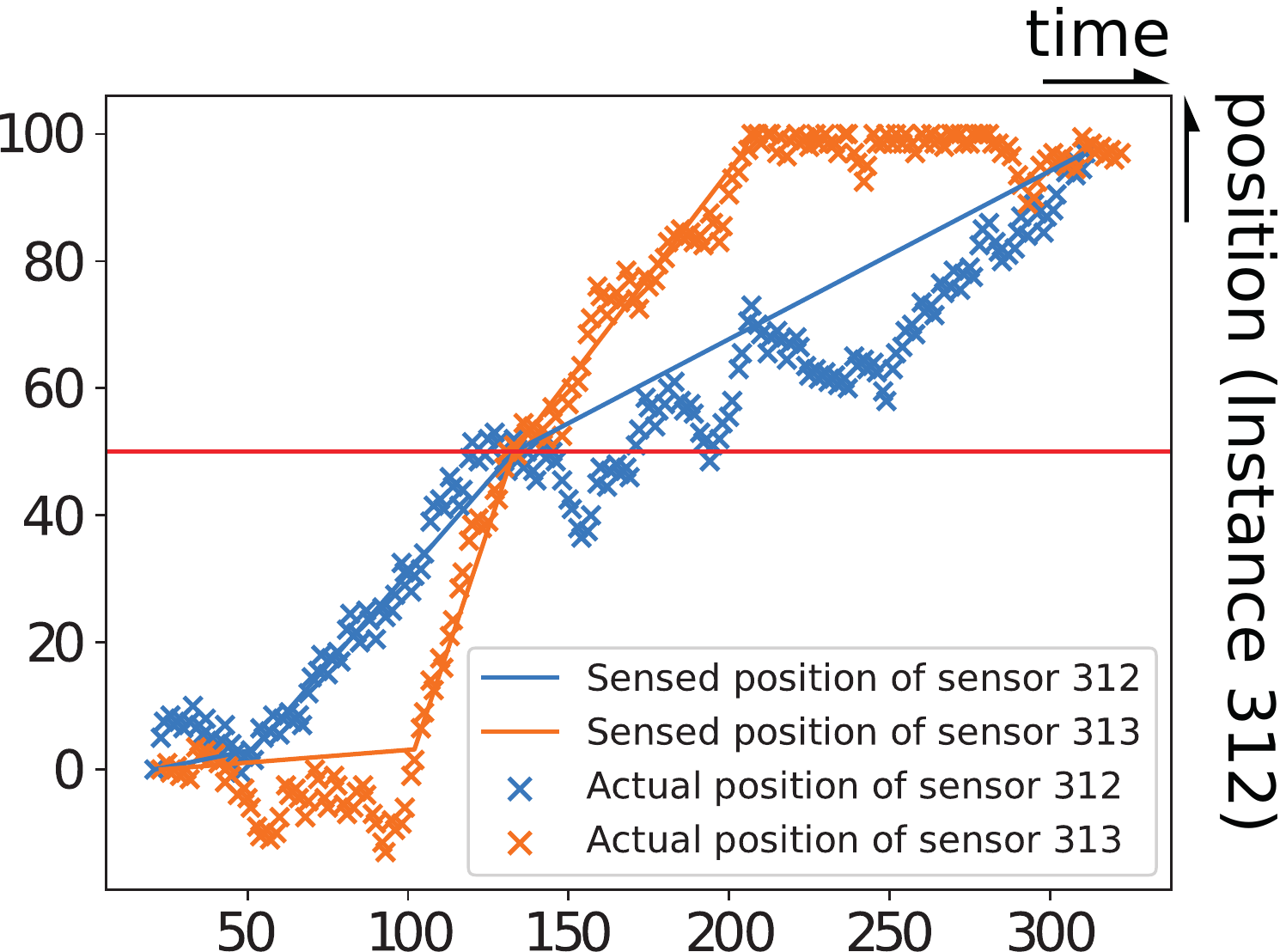}
        \caption{$\textrm{iRMSE} = 9.685$}
        \caption*{rectification enabled}
        \label{fig:rct-y2}
    \end{subfigure}
    ~
    \begin{subfigure}[t]{0.37\columnwidth}
        \includegraphics[width=\textwidth]{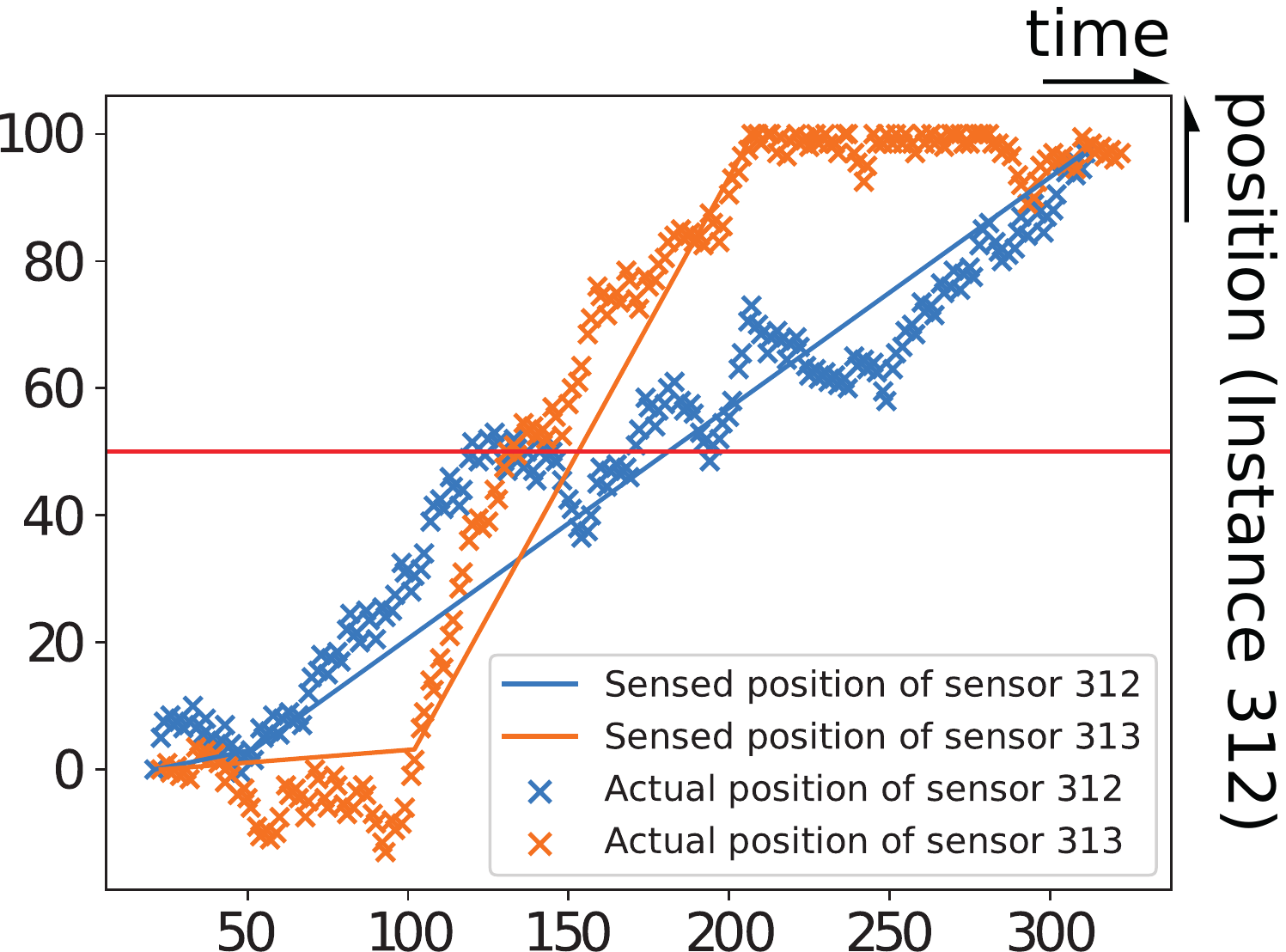}
        \caption{$\textrm{iRMSE} = 11.889$}
        \caption*{rectification disabled}
        \label{fig:none-y2}
    \end{subfigure}
    \caption{Comparison between enabled and disabled path rectification}\label{fig:y2-examples}
\end{figure}

Finally, we examine an instance in which path rectification is beneficial:
In Fig.~\ref{fig:y2-examples}, two nodes have different speeds before and after the junction where they first meet (red line).
They also, importantly, encounter each other before the final gateway. 
At the confluence point, new epochs are created for both nodes, permitting more accurate localization.
Path rectification only moves the position up to the first vertex where the other node could have possibly been encountered; therefore, it is not possible to `overshoot' the correct position.

\textbf{Scenario 4} In this final scenario (Fig.~\ref{fig:scenario-4}), we implemented a more realistic environment with longer paths from source to sink.
There are five nodes in this scenario.
The experiments show that GRAL with path rectification performs the best here ($\textrm{dRMSE}_P = 10.48$) while using checkpoints ($\textrm{dRMSE}_C = 11.72$) is outperformed by even the baseline algorithm ($\textrm{dRMSE}_B = 11.55$).
This demonstrates that, while checkpoints may be beneficial in environments with good gateway coverage, they may severely impact the result if gateways are sparsely placed.
The average localization error for the best configuration is $7.76\%$.

\section{Conclusion}\label{sec:conclusion}

In this paper, we introduce the~\gls{gral} algorithm for floating wireless sensor networks in pipe networks with a tree-like structure and sparse gateway coverage.
Our solution does not need auxiliary positioning systems like GPS, nor does it burden the sensors with any additional computation.
Instead, location estimates are computed for every measurement package once the node encounters a gateway and transmits them to a centralized backend.
In contrast to other range-free approaches, GRAL can deliver location estimates for moments where a node has no contact to any other node.
An important use case for our system is in vivo monitoring campaigns in pipe networks to detect leakages and infiltrations.

We evaluate~\gls{gral} and the two proposed extensions in a simulated environment with noisy node movement.
As a baseline for comparison, we use simple linear interpolation between the encountered gateways.
The results show that~\gls{gral} consistently outperforms the baseline and achieves a measurement localization accuracy between 4.81\% and 7.76\%, depending on the distance between consecutive gateways in the network and flow speed variability.
Using two extensions,~\emph{checkpointing} and~\emph{path rectification} in settings with multiple nodes, a sparse deployment of gateways, and a frequently confluenting pipe network, can improve estimation accuracy (RMSE = 9.89) over vanilla~\gls{gral} (RMSE = 10.03).
The advantage of \emph{path rectification} increases for networks with more junctions and longer pipe segments.

In the future, we plan to use floating sensor node prototypes to assess how well our solution is working in the real world.
It would also be promising to extend the algorithm further by taking flow characteristics of different parts of the pipe network (e.g. pipe width, materials) into account to further segment and adjust the position predictions.

\section*{Acknowledgments}

This work has been supported through grants by the German Ministry for Education and Research (BMBF) as WaterGridSense 4.0 (funding mark 02WIK1475D).

{
    \renewcommand*{\bibfont}{\footnotesize}
    \printbibliography
}

\end{document}